\documentclass[%
 reprint,
 amsmath,amssymb,
 aps,
]{revtex4-2}

\usepackage{graphicx}
\usepackage{dcolumn}
\usepackage{bm}
\usepackage{gensymb}  
\usepackage{float}
\usepackage{afterpage}
\usepackage{placeins}
\usepackage{hyperref}

\usepackage{xcolor}
\begin{document}

\preprint{APS/123-QED}

\title{Nontrivial topological phases in ``Zig-Zag'' arrays of polarization transmons}

\author{Konopleva E.E.}%
\affiliation{Center for Engineering Physics,
Moscow 121205, Russia}
\author{Fedorov G.P.}
\affiliation{Moscow Center for Advanced Studies
Moscow 123592, Russia}
\affiliation{Kotel’nikov Institute of Radio Engineering and Electronics
Russian Academy of Sciences
Moscow 125009, Russia}
\email{gleb.fedorov@phystech.edu}

\author{Astafiev O.V.}
\affiliation{Center for Engineering Physics,
Moscow 121205, Russia}
\affiliation{Moscow Center for Advanced Studies
Moscow 123592, Russia}
%


\date{\today}

\begin{abstract}
In recent years, quantum simulators of topological models have been extensively studied across a variety of platforms and regimes. A new promising research direction makes use of meta-atoms with multiple intrinsic degrees of freedom, which to date have been predominantly studied in the classical regime. Here, we propose a superconducting quantum simulator to study an extension of the well-known ``Zig-Zag'' model with long-range cross-polarization couplings using polarization transmons hosting degenerate dipole orbitals. We map the phase transitions of the extended ``Zig-Zag'' model both numerically and analytically using inverse participation ratios and topological invariants. We demonstrate the existence of in-gap localized trivial and Tamm edge states. With linearized meta-atoms, we show via electromagnetic modeling that the proposed arrangement closely reproduces the extended ``Zig-Zag'' model. This work paves the way towards experimental investigation of the previously inaccessible topological quantum many-body phenomena.

\end{abstract}

\maketitle


\section{\label{sec:level1} Introduction}

The study of topological phenomena and the consequent work on the development of supporting formalism was set in motion in the 1980's on the back of the discovery of the Quantum Hall Effect by Klitzing et al. \cite{Klitzing1980}. In the years that followed, much work has been done in expanding the field. In particular, with the search for intrinsically topological materials that possess topological phases in the absence of external magnetic fields, culminating in the proposal of the Quantum Spin Hall Effect by Kane and Mele \cite{Kane_2005, KaneMele2005}. The theory was then quickly followed by the experiments, which highlighted the difficulties of studying topology in natural materials -- from the difficult parameter control to the need to synthesize exceptionally pure materials in order to limit bulk state interference \cite{Checkelsky_2009, Butch_2010, Hasan_2010}. 

The limitations of natural materials and the complexities of classical simulation of large quantum systems lead researchers to employ analogue simulation in order to study these exotic condensed matter phases \cite{Feynman1982, Britton2012}. The use of controllable classical and quantum simulators enables precise tuning of system parameters, engineering of previously inaccessible physical regimes, and direct reconstruction of wavefunctions \cite{Britton2012, Olekhno2020}. The past decade has been marked by a proliferation of experimental realizations of topological models on a variety of platforms, ranging from several nanometers to centimeters in size, including quantum dots \cite{Kiczynski_2022}, isolated Rydberg atoms \cite{Leseleuc2019_Rydbergatoms}, cold atom condensates \cite{Atala2013_coldatoms, Jotzu2014_ultracoldatoms}, mechanical \cite{Susstrunck2015_mechanical, Lin_2021}, topoelectrical \cite{Lee2018_topoelectrical}, acoustic \cite{He2016_acoustic}, thermal \cite{Hu2022_thermal} and photonic devices \cite{Wang2009_ferriterods,Hafezi2013_siliconphotonics, Bleckmann2017_plasmonicwaveguide, kremer2020square}.

A particularly interesting new direction of recent studies explores artificial topological structures, the elementary cells of which have multiple intrinsic degrees of freedom (DOF), for instance, degenerate orbitals or oscillatory eigenmodes \cite{Savelev2020, Gao2023}. In the quantum regime, they are promising candidates for scaling towards classically intractable sizes, which would enable practical quantum simulation \cite{fedorov2021photon}. In this work, we focus on the ``Zig-Zag'' model, also known as the orbital SSH model, which is an extension of the prototypical SSH model with an additional degree of freedom -- electromagnetic field polarization. The ``Zig-Zag'' model serves as a photonic analogue to the Kitaev's chain \cite{leumer2020exact} with polarization-dependent nearest-neighbor electromagnetic interactions, first introduced in \cite{Poddubny2014}. Experimental studies of the model have been reported in arrays of dielectric and plasmonic nanoparticles, acoustic resonator chains and polariton micropillars \cite{Slob_2015, Sinev_2015, slobozhanyuk2016enhanced, Moritake_2022, Gao2023, stJean2017}. However, none of the studies have demonstrated quantum phenomena such as many-particle localization, predicted by the theory \cite{gorlach2017topological}, which is key for the intractability. The translation of the photonic ``Zig-Zag'' model onto the superconducting cQED platform, which could have facilitated such experiments, has not yet been done due to the absence of suitable multi-mode meta-atoms.

In this work, we propose and study a superconducting circuit architecture to simulate the extended ``Zig-Zag'' model in the quantum regime, leveraging the inherent scalability of mesoscopic superconducting circuits, the individual addressability of artificial atoms, and a diversity in the choice of topologies \cite{besedin2021topological}. We find that our circuit realizes an extended (long-range) ``Zig-Zag'' model, in which the next-nearest neighbour couplings are taken into consideration, similar to the extended SSH models studied before \cite{perez2019interplay}. We study its topological invariants and phases across the parameter space, demonstrating the emergence of in-gap localized states in both trivial and non-trivial regimes. Using electromagnetic simulation, we demonstrate the experimental feasibility of the design, with the simulation of the linearized polarization transmons showing good correspondence with the proposed extended ``Zig-Zag'' model. By laying the groundwork for experimental realization of the ``Zig-Zag'' model in a superconducting cQED platfrom, our work opens new avenues for further exploration of topological and many-body phenomena in the quantum systems with multiple orbitals. 


\section{Polarization transmon}

\begin{figure}
    \centering
    \includegraphics[width=0.95\linewidth]{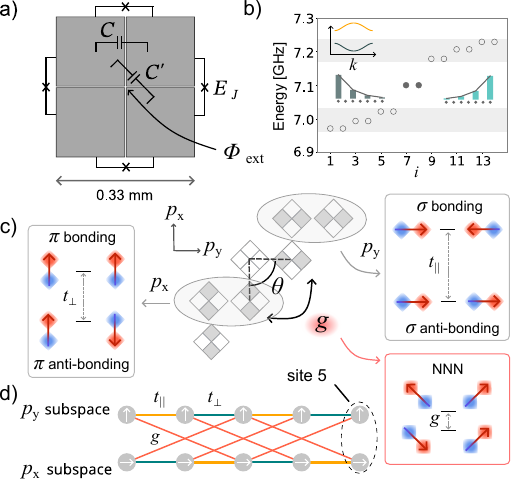}
    \caption{Polarization transmon as an elementary cell for the quantum topological ``Zig-Zag'' model simulator. \textbf{a)} A realistic geometry for an electrical circuit of the polarization transmon. Four closely positioned superconducting islands form axial and diagonal capacitors ($C, C^\prime \approx 33.9, 7.7 $ fF assuming $\epsilon_\text{eff}\approx6$). The nearest neighbors are connected by identical Josephson junctions (characterized by the $E_J/h \approx 14.9$ GHz constant) forming the quantum potential of the meta-atom. \textbf{b)} The doubly degenerate energy spectrum of a 7-site ``Zig-Zag'' chain with $\theta = 90 \degree$ and only nearest-neighbour couplings ($g=0$). Two pairs of degenerate bulk bands and two edge state distributions are also shown. \textbf{c)} The nearest-neighbour $\pi$-type coupling ($t_{\bot}$)  and $\sigma$-type coupling ($t_{||}$). Additional NNN cross-polarization coupling $g$ couples the $p_x$ and $p_y$ subspaces. \textbf{d)} Superconducting ``Zig-Zag'' model visualized as coupled $p_x$ and $p_y$ bosonic subspaces (orthogonal polarizations in the classical limit), which are SSH-chains with alternating couplings $t_{||}$ and $t_{\bot}$ coupled via $g$ interaction.}
    \label{fig:1}
\end{figure}

Previous work has investigated a new artificial atom design analogous to the nanoparticle dipoles and other multiple DOF meta-structures -- the polarization transmon, shown schematically in Fig.\ref{fig:1} a) \cite{radiophysics}. Deriving from the diamond-shaped artificial atoms, Josephson rhombi and other known quantum circuits, polarization transmon scheme is composed of 4 square-shaped superconducting islands connected via Josephson junctions \cite{Koch_2007, Orlando_1999, Dassonneville_2020, Diniz_2013, Doucot_2002}. The Hamiltonian of the polarization transmon for zero external flux is given by
\begin{equation}\label{Hamiltonian}
\begin{aligned}
    \mathcal H = \frac{4e^2}{2} &\left (\frac{\tilde{n}_1^2 + \tilde{n}_2^2}{4(C + C')} + \frac{\tilde{n}_3^2}{4C} \right) - \\
    &4E_J\cos(\tilde{\varphi}_1)\cos(\tilde{\varphi}_2)\cos(\tilde{\varphi}_3),
\end{aligned}
\end{equation}
where $\tilde \varphi_i$ and $\tilde n_i$ denote the generalized coordinates and momenta, and $e$ is the elementary charge. 

The behavior of the artificial atom near $\Phi_\text{ext} = 0$ resembles that of a standard transmon with the potential having a well-defined single well form repeated on a 3-dimensional lattice. Localized orbitals are formed in each isolated well, with the three lowest orbitals being oscillator-like and corresponding to two dipole and one quadrupole normal modes. We note that when $\Phi_\text{ext}$ is increased, the potential gradually deforms and bifurcates from a single-well to a symmetric double-well structure at $\Phi_\text{ext}/\Phi_0 = n \pm 0.5$, where the system resembles a flux qubit rather than a transmon. This regime is not well suited for our objectives, so we will assume that the external flux is tuned to zero for every site by an array of local bias coils, so that \eqref{Hamiltonian} is appropriate.

As could be inferred from the kinetic part of Eq. \ref{Hamiltonian}, this type of artificial atom exhibits two degenerate orthogonal dipole modes with fundamental frequencies $h f _{1,2} \approx \sqrt{16 E_J \tilde{E}_C}$, where $\tilde{E}_C = e^2/2(2C + C')$. The degenerate orbitals allow one to realize intersite couplings similar to the interactions between plasmonic nanoparticles (in the linear regime) or polariton micropillars (in the quantum regime).  

Despite concerns about the possible dephasing effects caused by the Aharonov-Casher effect in looped identical-junction circuits \cite{friedman2002aharonov}, we find that just as for the transmon, at $\Phi_\text{ext} =0$ the charge sensitivity is strongly suppressed for $E_J/\tilde{E}_C$ ratios characteristic of a physical circuit shown in Fig. \ref{fig:1} a) (see Supporting Information for further details). 

\section{Extended ``Zig-Zag'' model}
We begin the analysis by considering the ``Zig-Zag'' model with nearest-neighbour interactions only. In such zig-zag arrays, dipole modes of individual sites can be oriented perpendicularly or in parallel to the bond link, which leads to alternating coupling strength between the nearest neighbours (NN). Fig. \ref{fig:1} c) shows that the parallel alignment ($\sigma$ bonding) lowers the frequency of the symmetric oscillatory mode and results in the larger coupling energy $t_{||}$; conversely, the perpendicular alignment ($\pi$ bonding) lowers the anti-symmetric mode and results in the weaker coupling $t_{\bot}$. 


The quasimomentum-representation Hamiltonian of a ``Zig-Zag'' chain with the bond angle $\theta$ is given by \cite{Slob_2015, Lin2020}
\begin{equation}
\begin{aligned}
    \mathcal H_K^\text{ZZ} = & \begin{pmatrix}
        0 & \mathcal Q^\text{ZZ}_K  \\
        (\mathcal Q_K^\text{ZZ})^\dagger & 0
    \end{pmatrix}, \\
    \mathcal Q_K^\text{ZZ} = \overline{t}(1+e^{-iK}) \sigma_0 + &\Delta(1 +e^{-iK}\cos(\theta))\sigma_z + \\ 
    \Delta e^{-iK}& \sin(\theta)\sigma_x,
\end{aligned}
\label{eq:Bulkzigzag}
\end{equation}
where $2\Delta = t_{||} - t_{\bot}$, $2\overline{t} = t_{||} + t_{\bot}$, and $K$ is the crystal momentum.


``Zig-Zag'' model with both $\theta = 90 \degree$ and only NN couplings has a special symmetry, owing to which the dipole interactions lead to the formation of two independent polarization subspaces $p_{x,y}$ that are described by separate SSH chains with opposite staggering patterns. The doubly degenerate energy dispersion in this case is given by Eq. \ref{eq:ssh}, which indeed corresponds to the standard SSH model, see Fig. \ref{fig:1} b).
\begin{equation}\label{eq:ssh}
        E_{1-4}^\text{ZZ}(K) =  \pm \sqrt{t^2_{||} + t^2_{\bot} + 2t_{||}t_{\bot}\cos\left(K\right)  }.
\end{equation}

However, in our system, long-range dipole-dipole interactions manifest in the non-negligible cross-polarization NNN (next-nearest-neighbor) couplings with strength $g$, shown in Fig. \ref{fig:1} c) alongside the $\sigma$- and $\pi$-type NN couplings. In addition, due to the $\theta=90\degree$ arrangement, we find the co-polarized NNN couplings to be negligible and thus exclude them from the following analysis. Importantly, the effect of the cross-polarization interaction is to couple the $p_x$ and $p_y$ subspaces (as shown in Fig.\ref{fig:1} d), lifting the degeneracy of the energy spectrum in both the bulk and at the edge. 

To take this effect into account, we introduce a modified bulk Hamiltonian with NNN couplings where a total of 4 sites per unit cell (2 per polarization subspace) is contained, and that is given by
\begin{equation}
     \mathcal H_K = 
    \begin{pmatrix}
        0 & d(K) & 2g\cos(K) & 0 \\ 
        d^\dagger (K) & 0 & 0 & 2g\cos(K) \\
        2g\cos(K) & 0 & 0 & d'(K) \\
        0 & 2g\cos(K) & d'^{\dagger}(K) & 0 \
    \end{pmatrix}, 
\end{equation}
where $d(K) = t_{||} + t_{\bot}e^{iK}$, $d'(K) = t_{\bot} + t_{||}e^{iK}$ (refer to Fig. \ref{fig:1} d) for the schematic representation). 

Alternatively, using the Pauli matrices, we can write the Hamiltonian in a shorter form:
\begin{equation}
    \begin{aligned}
    \mathcal H(K) = \text{Re}[d_{+}] \tau_0 \sigma_x + \text{Im}[d_+] \tau_0\sigma_y + \text{Re}[d_{-}]\tau_z \sigma_x + \\ \text{Im}[d_{-}] \tau_z \sigma_y + 2g\cos{(K)} \tau_x \sigma_0,
    \end{aligned}
\end{equation}
where $d_{\pm} = (d(K) \pm d'(K))/2$. It is easy to check that this Hamiltonian is chiral with the chiral symmetry operator $\Gamma = \tau_z \otimes \sigma_z = \text{diag} ({1, -1, -1, 1})$, indicating that one can rewrite the matrix in the off-diagonal form and compute the topological invariants of this system. The off-diagonal form is obtained by performing a basis transformation $(A_x, B_x, A_y, B_y) \rightarrow (A_x, B_y, B_x, A_y)$, resulting in 
\begin{equation}
    \mathcal H_K = \begin{pmatrix}
        0 & \mathcal Q_K \\
        \mathcal Q^\dagger_K & 0 \\
    \end{pmatrix}
\end{equation}
with 
\begin{equation}
\begin{aligned}
    \mathcal Q_K =  \left( \overline{t} + \frac{t_{\bot}e^{iK} + t_{||}e^{-iK}}{2}\right) \sigma_0 + \\ \left ( \Delta  + \frac{t_{\bot}e^{iK} - t_{||}e^{-iK}}{2} \right) \sigma_z + 2g\cos(K)\sigma_x.
\end{aligned}
\end{equation}

Note that this form is achieved in a different basis compared to Eq. \eqref{eq:Bulkzigzag}. The energy bands can be calculated analytically for this Hamiltonian by diagonalizing the $\mathcal Q_K^\dagger \mathcal Q_K$ matrix \cite{Slob_2015}.

\begin{figure}
    \centering\includegraphics[width=1\linewidth]{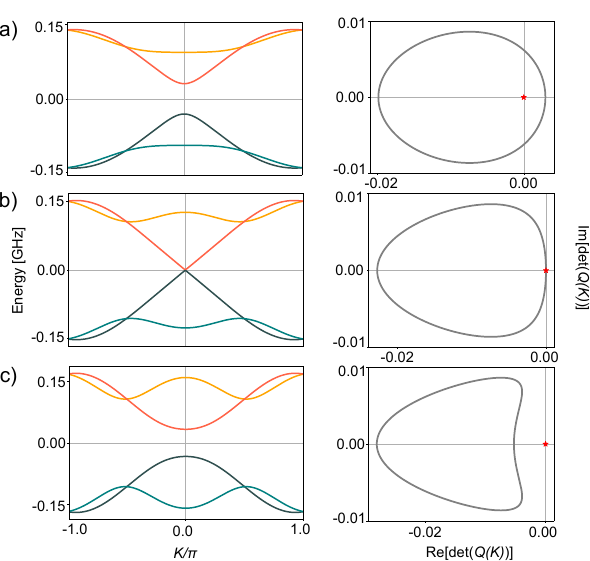}
    \caption{Extended ``Zig-Zag'' band structure $E_{1-4}(K)$ (left column) and winding diagram (right column) for $t_{||} = -0.1$, $t_{\bot} = 0.037$ and varying NNN coupling strength.\textbf{ a)} $g = 0.016$: the system is gapped and $\mathcal{W} = 1$. \textbf{b)} $g = 0.0315$: the system is gapless and $\mathcal{W}$ is undefined. \textbf{c)} $g = 0.048$: the system is gapped and $\mathcal{W} = 0$.   }
    \label{fig:topinv}
\end{figure}

\begin{figure*}[t]
    \centering
    \includegraphics[width=\linewidth]{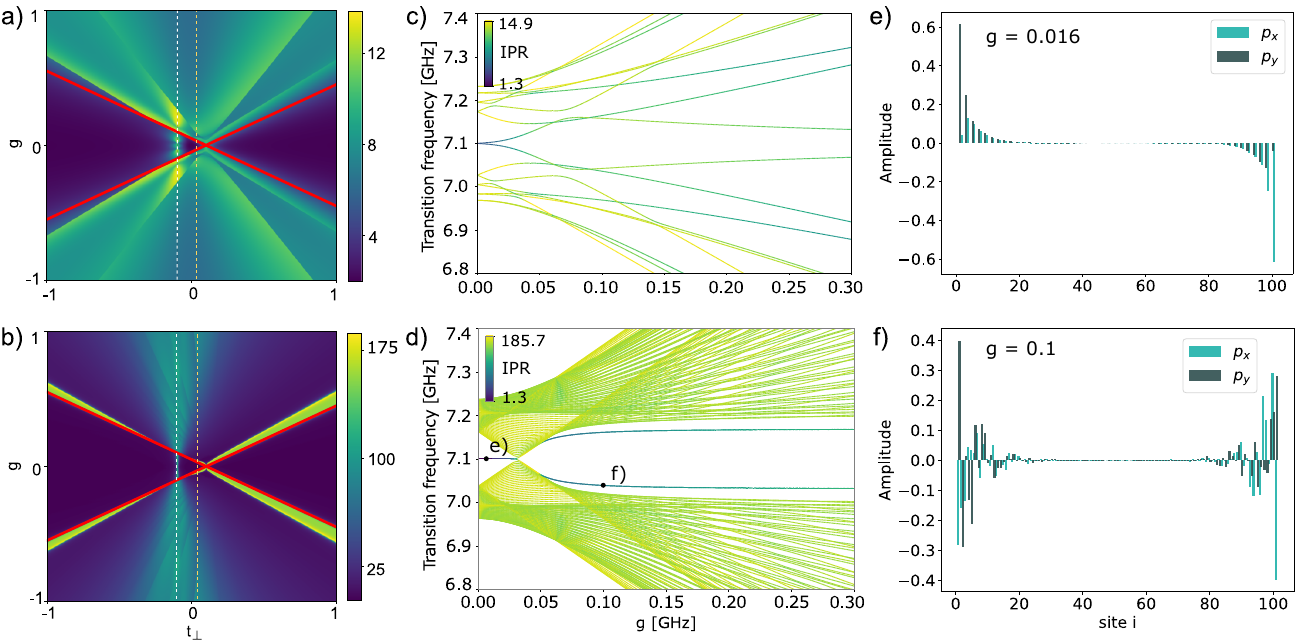}
    \caption{Phase (IPR) and transition frequency diagrams for extended ``Zig-Zag'' chains of length $N=9$ \textbf{a,c)} and 101 \textbf{b,d)}. In \textbf{a,b)}, the white dashed line (left) indicates $t_\bot = t_{||}$, the red lines are analytic phase transition points and the orange dashed line (left) indicates $t_\bot = 0.037$, where the spectra c,d) are calculated. \textbf{e,f)} The mid-spectrum eigenstates for $g = 0.016$ and $g = 0.1$ respectively. As one can see in the top panel, both polarizations contribute to the topological edge state due to the NNN interaction.}
    \label{fig:phasediagram}
\end{figure*}

\subsection{Topological invariants}
The extended chiral symmetric ``Zig-Zag'' model exhibits topologically distinct phases which are characterized by the winding number \cite{sato_topology_2011, Slob_2015}:
\begin{equation}
\begin{aligned}
    \mathcal{W} =  \frac{i}{2\pi} & \int^\pi_{-\pi} dK \frac{d \text{ ln}   \text{ det} [\mathcal Q_K]}{dK} = \\ -\frac{1}{2\pi} &\int_\mathcal{C} d \text{ arg \{det}[\mathcal Q_K]\},
\end{aligned}
\end{equation}
where $\mathcal{C}$ denotes the contour swept by $[\mathcal Q_K]$ over the first Brillouin zone.
The topological invariant counts the number of times the Hamiltonian's vector in the Bloch vector space encircles the origin, providing a direct link between the system's bulk geometry and the presence of protected boundary states (the bulk-boundary correspondence). Fig. \ref{fig:topinv} shows a phase transition between topologically non-trivial and trivial states guaranteed by the Hamiltonian's chiral symmetry. In Fig. \ref{fig:topinv} a) the spectrum is gapped and the corresponding winding diagram indicates $\mathcal{W} = 1$. Then as $g$ is increased, the system passes through a gapless conducting state for which $\mathcal{W}$ is undefined. Further increase in $g$ results in a gapped bulk with $\mathcal{W} = 0$. This observation is consistent with the behavior of other well-known prototypical 1-dimensional topological models.

A more general condition for a phase transition is derived analytically by enforcing that both $\text{ Im [det} [\mathcal Q(K)]]$ and $\text{ Re [det} [\mathcal Q(K)]]$ are equal to zero for some $K$. The former condition is satisfied if either $t_{||} = \pm t_{\bot}$ or $\sin{(K)} = 0$. The solution to this set of equations is $\{K = 0\} \land \{(t_{\bot} + t_{||})^2 = 4g^2\}$. Transitioning across these lines results in a change of $\mathcal{W}$ from 1 to 0 or vice versa. Furthermore, there exists a transition between topologically indistinguishable phases ($ 0\leftrightarrow 0$ or $ 1\leftrightarrow 1$) accompanied by the band gap closure at $t_{||} = t_{\bot}$ and at $\{t_{||} = -t_{\bot}\} \land \{g=0\}$. We study these transitions in greater detail with the help of numerical diagonalization in the following section.

\begin{figure}[tp]
    \centering
    \includegraphics[width=\linewidth]{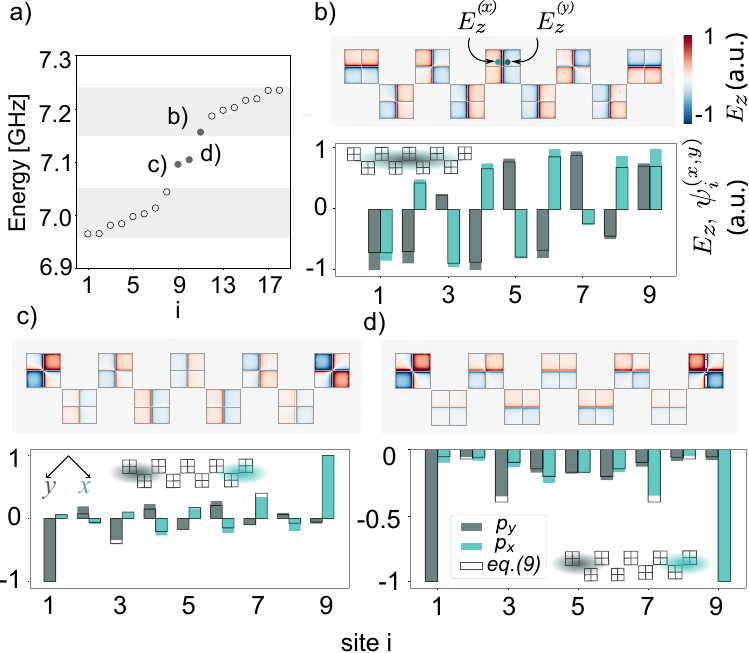}
    \caption{A quantitative comparison between three classical electromagnetic eigenmodes and single-excitation wavefunctions calculated from the second quantization formalism for the proposed architecture, $N=9$.  \textbf{a)} Energy spectrum obtained from the second quantization Hamiltonian. \textbf{b)} Lowest mode of the higher bulk band. \textbf{c)} Lower frequency edge mode. \textbf{d)} Higher frequency edge mode. In the top parts of the b-d) panels, the data from the FEM modeling are shown; the vertical component of the electric field directly above the electrode plane is plotted with color. In the bottom, extracted amplitudes of perpendicular polarization components $p_{x,y}$ are shown with bars, and the amplitudes obtained from second quantization Hamiltonian are shown in black outline.}
    \label{fig:hfss}
\end{figure}

\subsection{Phase Diagrams}

To plot a phase diagram of the extended ``Zig-Zag'' model we numerically diagonalize a finite chain given by the second quantization model:
\begin{equation}\label{eq:ZigzagNNN}
    \begin{split}
        \mathcal H & = \mathcal H_{t_{||}, t_{\bot}}^\text{SSH} + \mathcal H_{t_{\bot}, t_{||}}^\text{SSH} + \mathcal H_\text{g}, \\[10pt]
        \mathcal H_{t_{||}, t_{\bot}}^\text{SSH} & = \sum^{N-1}_{i=1} \left\{ 
        \begin{array}{ll}
            t_{||} c^\dagger_{i}c_{i+1} & (\text{odd}\  i) \\
            t_{\bot} c^\dagger_{i}c_{i+1}& (\text{even } i)
        \end{array} \right.  + \text{h.c.}, \\
        \mathcal H_{t_{\bot}, t_{||}}^\text{SSH} &= \sum^{N-1}_{i=1} \left\{ 
        \begin{array}{ll}
            t_{\bot} d^\dagger_{i}d_{i+1} & (\text{odd } i) \\
            t_{||} d^\dagger_{i}d_{i+1} &( \text{even } i)
        \end{array} \right.  + \text{h.c.}, \\
        \mathcal H_\text{g}& = \sum ^{N-2}_{i=1} g (d^\dagger_{i}c_{i+2} + d^\dagger_{i+2}c_{i}) + \text{h.c.}, \\
    \end{split}
\end{equation}
where $c_i^\dagger\ (d_i^\dagger)$ is the creation operator on the $i$-th site of the $p_{x(y)}$ subspace.

For a set of parameters, we calculate the inverse participation ratio (IPR) of the edge modes. It is given by $\text{IPR} = \big(\sum_{i,\alpha=(x,y)} |\psi_i^{(\alpha)}|^4 \big)^{-1}$, and is a measure of the state’s localization with the minimum of unity for single-site wavefunctions. Fig. \ref{fig:phasediagram} a,b) show a cone-like phase diagram for a mid-spectrum state of a 9-site and a 101-site chains, where the dark blue colour corresponds to a high degree of localization, and green corresponds to the de-localized state. The phase diagrams are asymmetric with respect to the centreline $t_{\bot} = 0 $ due to the presence of the $t_{||}$ coupling, the magnitude of which is indicated by the white dashed line $t_{\bot} = t_{||}$. The red lines depict the analytically derived phase transition condition from the previous section; as the number of sites in the chain is increased to 101, the IPR phase diagram converges to show good agreement with the phase transitions predicted analytically. 

In Fig. 3 c,d) we plot the transition frequencies (discrete bands) calculated along the orange dashed line from Fig. \ref{fig:phasediagram} a) and indeed observe the expected phase transition, where at $g \approx 0.06$ the edge modes merge with the bulk. Notably, with sufficient chain length, \textit{two} pairs of chiral defect edge states re-emerge inside the band gap at high $g$s. 

In contrast to Fig. \ref{fig:1} b), where the NN ``Zig-Zag'' spectrum is shown, the topological edge modes of the extended ``Zig-Zag'' model always come in doublets -- one symmetric mode and the other antisymmetric, detuned in frequency from each other -- for any number of sites. For the odd-sited chains, the excitations localize as $p_x$ components on one end and as $p_y$ components on the other, while in the even-sited chains, the excitation localizes in the form of either $p_y$ or $p_x$ components on both ends. Fig. \ref{fig:phasediagram} e) shows an antisymmetric topological edge mode of a 101-site chain for $g = 0.016$, where different polarizations indeed localize at opposite ends. In the trivial region, Fig. \ref{fig:phasediagram} f) shows a defect (Tamm) edge mode, and we see that its spatial distribution has a disordered phase relationship between components and extends slightly past the topological ones into the bulk region. We also check that there is no topological protection for these states (see Supporting Information, Sec. III).

\section{Experimental feasibility}

To demonstrate the experimental feasibility of a polarization-transmon-based ``Zig-Zag'' simulator, we study arrays of polarization transmons in the linear limit (with the Josephson junctions replaced by linear inductors of comparable inductance) using finite element method (FEM) modeling. To verify the proposed implementation, we compare the results obtained via FEM and the numerical diagonalization of \eqref{eq:ZigzagNNN}.   

In Fig. \ref{fig:hfss} a) we plot the energy spectrum of a 9-site chain obtained from the second quantization Hamiltonian. Not that the edge modes are separated in energy due to the NNN coupling. Top rows of Fig. \ref{fig:hfss} b), c) and d) show charge distributions on the artificial atoms, proportional to the electric field components $E_z^{(x)} $ and $E_z^{(y)}$ perpendicular to the silicon substrate at points $x$ and $y$ on each site; the field projections are taken 0.1 $u$m above the $xy$ plane. Fig. \ref{fig:hfss} b) shows a bulk mode at the lower edge of the upper bulk band where we see a de-localized excitation with relative amplitudes and phases (bars) in good correspondence with the proposed theory (black outline). Fig. \ref{fig:hfss} c) and d) show the edge modes, in which, as predicted, the excitations localize at both ends in symmetric and anti-symmetric superpositions, with non-zero wavefunction components on each site of the chain. The findings demonstrate that the proposed design is experimentally viable and effectively captures the extended ``Zig-Zag'' model.

\section{Conclusion}\label{sec:conclusion}

In this work, we have proposed and studied a design for an orbital ``Zig-Zag'' model simulator based on a novel type of superconducting artificial atoms. We find that in our simulator, long-range couplings beyond the nearest neighbors play an important role. We also show that these NNN interactions are responsible for the coupling between polarization subspaces, which would be uncoupled in their absence. We derived the analytical model for this extended version of the ``Zig-Zag'' model in the periodic limit and demonstrated its chiral symmetry. The topological phases in the parameter space were mapped numerically using IPR and analytically by the winding number. 

Notably, for high enough NNN couplings $g$ (higher than are reached physically in our design) and for long enough chains, we observe the appearance of pairs of in-gap edge states. We distinguish the defect states from the topological ones based both on the value of the topological invariant and their immunity to disorder. 

Finally, we demonstrate in a preliminary 3D EM modeling (in the classical limit) that the extended ``Zig-Zag'' model is accurately emulated by a zig-zag chain of superconducting polarization transmons. This is supported by the comparison of the wavefunctions from the second quantization formalism and the eigenmodes obtained via FEM, which are in good agreement. 

This study lays the groundwork for the experimental demonstration of previously inaccessible quantum many-body phenomena on a new platform as well as enabling prospective  experiments with higher-order topological materials \cite{jacqmin2014direct, li2020higher}.

\begin{acknowledgments}
E.K. and G.F. wish to acknowledge the RSF grant no. 25-22-00280. We heartfully acknowledge M.A. Gorlach for his expertise and fruitful discussions.
\end{acknowledgments}


\bibliography{bibliography}

\end{document}